# Lattice correlation of Hubbard excitons in a Mott insulator Sr$_2$IrO$_4$ and reconstruction of their hopping dynamics via time-dependent coherence analysis of the Bragg diffraction

Yuelin Li[1*], Richard Schaller[2], Mengze Zhu[3], Donald A. Walko[1], Jungho Kim[1], Xianglin Ke[3], Ludi Miao[4], and Z. Q. Mao[4]

[1]Advanced Photon Source, Argonne National Laboratory, Argonne, Illinois 60439, USA

[2]Center of Nanoscale Materials, Argonne National Laboratory, Argonne, Illinois 60439, USA

[3]Department of Physics and Astronomy, Michigan State University, East Lansing, MI 48824, USA

[4]Department of Physics and Engineering Physics, Tulane University, New Orleans, LA 70118, USA

In correlated oxides the coupling of quasiparticles to other degrees of freedom such as spin and lattice plays critical roles in the emergence of symmetry-breaking quantum ordered states such as high temperature superconductivity. We report a strong lattice coupling of photon induced Hubbard excitonic quasiparticles in spin-orbital coupling Mott insulator Sr$_2$IrO$_4$. Combining time-resolved optical spectroscopy techniques, we further reconstructed spatiotemporal map of the diffusion of quasiparticles via time-dependent coherence analysis of the x-ray Bragg diffraction peak. Due to the unique electronic configuration of the exciton, the strong lattice correlation is unexpected but extends the similarity between Sr$_2$IrO$_4$ and cuprates under highly non-equilibrium conditions. The coherence analysis method we developed may have important implications for characterizing the structure and carrier dynamics in a wider group of oxide heterostructures.

---

[*] ylli@aps.anl.gov

Spin-orbital coupling Mott insulator Sr$_2$IrO$_4$ (SIO) [1,2] shares with cuprates several distinctive features that are characteristics for high temperature superconductor (HTSC): quasi-two dimensional square lattice, single Hubbard Band insulator, spin 1/2[1], and Heisenberg antiferromagnetic coupling[3]. Upon electron doping, Sr$_2$IrO$_4$ also produces the Fermi arc parallel to that observed in HTSC cuprates[4]. The structure, spin, and electronic phase similarity between the two make SIO an ideal test bed for understanding the material properties essential for HTSC and lead to the speculation of superconductivity in iridium oxides upon doping[5,6].

A cardinal issue in HTSC is the electron paring mechanism. For conventional superconductors, electrons in Cooper pairs are bonded by lattice vibrations (i.e., phonons). For HTSC, although the high transition temperature and the unconventional d-wave pairing symmetry suggests that the pairing mechanism may be associated with strong electron-electron correlation (EEI) and/or spin fluctuations, evidence shows that electron–phonon interaction (EPI) may play an important role as well[7,8]. One way to interrogate the electron phonon coupling is via photon excited non-equilibrium quasiparticles[9,10] that relaxes via both EEI and EPI that leads to strong lattice correlations[11,12]. The strong lattice correlation of O-*p* to Cu-*d* excitation in the HTSC cuprate parent compound La$_2$CuO$_4$ (LCO) [11] has not been reproduced in any other materials.

Despite the similarities, SIO is a Mott-insulator while cuprates are charge transfer (CT) insulators with completely different electronic structure[5]. The active orbital in cuprate is the strongly anisotropic $e_g$ $d_{x^2-y^2}$, whereas in SIO the active orbital is an equal superposition of the $t_{2g}$ $d_{xy}$, $d_{xz}$, and $d_{yz}$ wave with less anisotropy. Although strong electron-phonon coupling has been suggested by optical spectroscopy in temperature dependent analysis in iridate[13–15], the difference in electronic structure leads to a different quasiparticle configuration as compared to cuprates[16] and alludes to different EPI. It is therefore instructive to explore if photo-doping in SIO also generate the strong lattice correlation similar to that observed in LCO.

We report a surprisingly strong lattice response of SIO thin films to optical excitation using time-resolved x-ray diffraction, which is directly correlated to electronic dynamics probed via transient optical absorption spectroscopy (TAS). The excitation photo energy dependence suggests that the quasiparticle is a long lived Hubbard exciton[16]. We reconstructed a spatiotemporal transport map of the excitons along the c-axis based on the dynamic structure response via coherence analysis of the X-ray diffraction peak.

The lattice dynamics is measured via observing the shift and broadening of the (0 0 12) crystallographic diffraction peak of (0 0 1) oriented SIO thin films. A schematic of the experiment is shown in Fig. 1 (a). An example for a 100 nm film at 1.5 eV excitation photon energy is shown in Fig. 1 (b), where the diffraction peak is shifted by 0.025 Reciprocal Lattice Units (RLU), corresponding to a strain of 0.21%, indicating a significant expansion of the c-axis which decays over a time of 20 ns (Fig. 1 (c)). Accompanying the shift, there is a significant broadening of the diffraction peak (Fig. 1(d) and (e)). Strikingly, the broadening is much smaller for the 1.5 eV than for the 3.0 eV excitation energy, whereas the peak shift is much larger (Figs. 1 (b, c) and S1). In fact, while the temporal recoveries of the strain are almost identical for the two excitation energies, the recoveries of the peak broadening are drastically different. For the 3.0 eV excitation, there is a rapid initial drop within the first 2 ns (Fig. 1 (d)), while for the 1.5 eV excitation case, the peak broadening increases initially to a maximum at about 1 ns followed by a slow decay. The peak strain is linearly dependent on the laser fluence (Fig. S1).

There is also a strong thickness dependence of recovery of the structure dynamics (Fig. 2 (a)): the thicker the film, the longer the recovery time. The strain as a function of time is found to be of a stretched exponential function $\varepsilon(t)=a+b\exp(-(t/\tau)^\beta)$, where $a$ is a long time scale strain decaying over more than 150 ns, $b$ is the amplitude at time zero, and $\beta$ is the stretched exponential, respectively. The 1/e recovery time $\tau$ extracted is 1.4±0.4, 8.1±1, and 15±2 ns, respectively for the 20, 50 and 100 nm films. The peak broadening $\Delta w$ is also strongly dependent on the film thickness (Fig. S1).

There are several commonly known causes for photon induced lattice expansion, including photostriction via the piezoelectric effect[17], the deformation potential [18], and heating. SIO is not piezoelectric so photostriction plays no role. The deformation potential of SIO is negative due to the negative $dE_g/dp$ [19,20], thus the effect causes lattice contraction and is inconsistent with our observation. The thermal expansion can also be excluded due to the insensitivity of the lattice parameter to the temperature [21] and the insensitivity of the photo induced strain to the sample temperature (Fig. S2).

In the TAS measurement, immediately after the laser excitation, a photon induced transparency, i.e., negative optical density (OD), is observed at about 1 eV (Figure S3) and is a manifestation of the overall electronic response of the system to the photo excitation. There is a fast and a slow component. The fast component lasts less than 1 ps and can be attributed to recombination and cooling of the photo excited carrier via phonon and magnon emission[13]. When convolved with a 100 ps gate to emulate the X-ray data resolution, the OD and the strain dynamics overlap with each other, exhibiting the same thickness dependence (Fig. 2 (a)).

We attribute the lattice effect to the presence of excited carriers, or quasiparticles. The thickness dependent dynamics in Fig. 2 thus implies that quasiparticles are long-lived and their dynamics is dominated by diffusion and recombination at the surface or interface [22]. Furthermore, the stretched exponential characteristics indicates that the hopping of the excitons is a continuous time random walk with time dependent diffusion constant [23,24]:

$$\frac{\partial}{\partial t}N(t,z) = \frac{\partial}{\partial z}\left(Dt^{-\gamma}\frac{\partial}{\partial z}N(t,z)\right), \qquad (1)$$

where $N$ is the quasiparticle density, $D(t)=Dt^{-\gamma}$ is diffusion parameter with $1>\gamma>0$, where $\gamma$ is a measure of the trap energy distribution[24]. The equation has the following initial and boundary conditions,

$$N(0,z) \propto \exp\left(-\frac{z}{\alpha}\right), \qquad (2)$$

$$\frac{\partial}{\partial z}N(t,z) = \frac{s(z)}{D(t)}N(t,z)\bigg|_{z=0 \text{ or } Z}. \qquad (3)$$

where $\alpha$ is the optical absorption length (30 and 70 nm for 3.0 and 1.5 eV photon energy [25], respectively), $z=0$ is the free surface of the film, $z=Z$ is the film/substrate interface, and $s$ is the exciton dissociation velocity at the surface.

The strain is proportional to the density of quasiparticles (using the linear excitation fluence dependence in Fig. S1). It is straight forward to see that the shift of the diffraction peak in the x-ray measurements correspond to the average phase or strain of the unit cells. The broadening of the diffraction peak, on the other hand, is the manifestation of the decoherence or increase of the phase/strain spread of the diffracting unit cells. Solving equations (1-3) for the 100 nm film by adjusting $s$, $D$ and $\gamma$ and then Fourier transforming the resultant spatiotemporal strain map (Fig. 3 (a, b)) to fit the strain (average phase) data in Fig. 1 (c), we quantitatively reproduced the broadening (phase spread) of the diffraction peak data, as shown in Fig. 1 (d, e). We also reproduced the fluence dependence of the strain and broadening for the 100 and 20 nm films (Fig. S1 (b)). The fitting parameters are summarized in Table S1.

Based on this diffusion model, data in Fig. 1 (c-d) can be interpreted as the result of the difference in the initial quasiparticle spatial distribution due to the different absorption lengths for the two excitation photon energies thus the different manifestation of the decoherence of the diffracting unit cells. For the 3.0 eV excitation case, the photon penetration depth is much shorter (30 nm), thus a smaller average quasiparticle density with a larger density deviation over the film thickness. This leads to a smaller average phase shift thus peak strain but larger total decoherence thus larger peak broadening, in contrast to the larger penetration depth of the 1.5 eV excitation (70 nm) thus the opposite behavior in strain and peak broadening. As $OD(t) \propto \int N(t,z)\,dz$ and $\varepsilon(t) \propto \int N(t,z)\,dz/Z$, this also explains the overlapping of the OD and the strain dynamics in Fig. 2. The validity of the 1D diffusion model indicate the localized nature of the exciton with respect to the c-axis and a relatively fast 'sharing' of the exciton in the a-b plane, consistent with the anisotropy of the hopping integral [5] and the conductivity[26], and is a topic worthy of further investigation. The annihilation of the quasiparticles at the surface/interface can be attributed to orbital reconstruction thus the change of band structure[27].

In LCO, the optical excitation leads to *p-d* CT excitons. By examining our data for excitation photon energy ranging from 0.5 to 3.0 eV (Fig. 4 (a)) and the band structure of SIO (Fig. 4 (b))[2,14], excitation occurs both over CT *p-d* (3 eV) and *d-d* transitions (< 1 eV). However, after normalizing at time zero, the strain recovery for all these excitation photon energies collapse into one curve for the 20 nm film (Fig. 4 (a)). Given that the 0.5 eV excitation can only excite the transition from the lower $J_{eff}=1/2$ state (Lower Hubbard band, LHB) to an unoccupied $J_{eff}=1/2$ state (upper Hubbard band, UHB)[28], we conclude that the quasiparticle is the Hubbard exciton[16], with an electron at the bottom of the UHB and a hole at the top of the LHB. The relaxation into this configuration in the first picoseconds can be via *p-d* and *d-d* orbital hybridization [13,29].

For LCO, the lattice distortion was explained by the modification of the cohesion energy due to net O-Cu CT arise from the *p-d* excitation[11]. This is not readily applicable to SIO as the Hubbard exciton does not involve a net *O-Ir* CT. Recently it is proposed that a *d-d* CT instability with lattice deformation may explain the lattice expansion in LCO [30]. In fact, CT instability with deformed lattice is also suggested to be a common cause for carrier localization in other transition metal oxides [31,32]. In this scenario, the *d-d* CT excitation leads to a metastable CT state accompanied by a lattice deformation that localizes the electron [30]. The difficulty of this model is that it predicts a saturation of c axis expansion at high laser fluence which was not observed in LCO and in current measurement for SIO in the range of the laser fluence used (up to about 0.4 photons per Ir site). Note that, our coherent analysis is effectively a scaled down version of the atomic resolution capable Coherent Bragg Rod Analysis (COBRA) [33], which may provide the unitcell structure correlates of the Hubbard excitons in a future experiment. Conceptually, excitation of Hubbard excitons modifies the competition between the spin-obit coupling and the on-site Coulomb interaction in $SIO^2$ which makes the electronic structure highly susceptible to external conditions including the temperature[14,15], pressure [19], and the epitaxial strain[34,35].

We demonstrate that SIO shares with LCO a strong EPI effect that persists under non-equilibrium conditions, further extending the role of SIO in understanding the mechanisms of HTSC[4]. Emerging evidence show that lattice vibration in cuprate plays also important role in competing phases including charge order[36], charge density waves[37], the interlayer coupling of electron transport[38], and pseudo gap formation[39]. Quasiparticle-lattice coupling similarities between SIO and cuprate under non-equilibrium states thus opens a new scope for searching properties essential for understanding the feasibility of unconventional superconductivity in SIO in a more holistic way. The method we used, i.e., time dependent coherence analysis of the diffraction peak, with refinement, may provide information of the dynamic unit cell structure necessary for theoretical understanding of the exciton phenomena in a wider class of strongly correlated materials.

**Methods**

The optical excitation, x-ray diffraction experiment was performed at Sector 7 at the Advanced Photon Source[40]. A laser beam with photoenergies ranging from 0.5 to 3.0 eV (wavelength from 0.4 to 2.5 μm) and pulse duration of 60 fs impinged on the sample, with polarization lies within the O-Ir plane. The 3.0 eV pulse is obtained by frequency doubling the 1.5 eV fundamental energy of a 2.5 W Ti: Sapphire laser system. Pulses at 0.95 and 0.5 eV are generated using an optical parametric amplifier. The laser spot, about 0.5-0.9 mm diameter, always overfilled the x-ray footprint, which was about 50 μm. The delay between the x-ray and the laser was adjusted electronically. The temporal resolution was limited by the x-ray pulse duration of about 100 ps. An avalanche photo diode was used as the detector.

The transient absorption spectroscopy (TAS) experiment with photoenergy of 1.5 and 3.0 eV and broadband probing for energy from about 0.8 to 1.4 eV were performed at the Center of Nanoscale Materials at Argonne National Laboratory. The 3.0 eV pulse was obtained by frequency doubling of the 1.5 eV fundamental output of a 2 kHz Ti: Sapphire laser system. The probe pulse was generated by focusing a portion of the 1.5 eV fundamental into a 13-mm thick sapphire crystal. A mechanical delay stage controlled excitation-probe time-delay, and excitation pulses were alternately blocked for calculation of transient signals. The excitation spot size was 0.38 mm and the probe pulse was 0.15 mm in diameter, respectively.

Epitaxial thin film samples of (0 0 1) SIO were grown on (0 0 1) SrTiO$_3$ using the pulsed laser deposition (PLD) method with a KrF excimer laser ($\lambda$ = 248 nm). A stoichiometric SIO polycrystalline pellet was used as the target. During deposition, the substrates were kept at 1080º C with oxygen partial pressure $p_{O2}$ = 150 mTorr, in a vacuum chamber with a base pressure of $10^{-4}$ mTorr.

**Acknowledgement**

This research used resources of the Advanced Photon Source, a U.S. Department of Energy (DOE) Office of Science User Facility operated for the DOE Office of Science by Argonne National Laboratory under Contract No. DE-AC02-06CH11357. M. Z and X. K. acknowledges the start-up funds from Michigan State University. Work at Tulane University was supported by the NSF under Grant DMR-1205469 and DOD ARO under Grant No. W911NF0910530.

**Contributions**

YL conceived the experiment. YL, RS, MZ and DW performed the experiment. LM and ZQM grew the samples. All authors contribute to the writing of the manuscript.

**Competing financial interests**

The authors declare no competing financial interests.

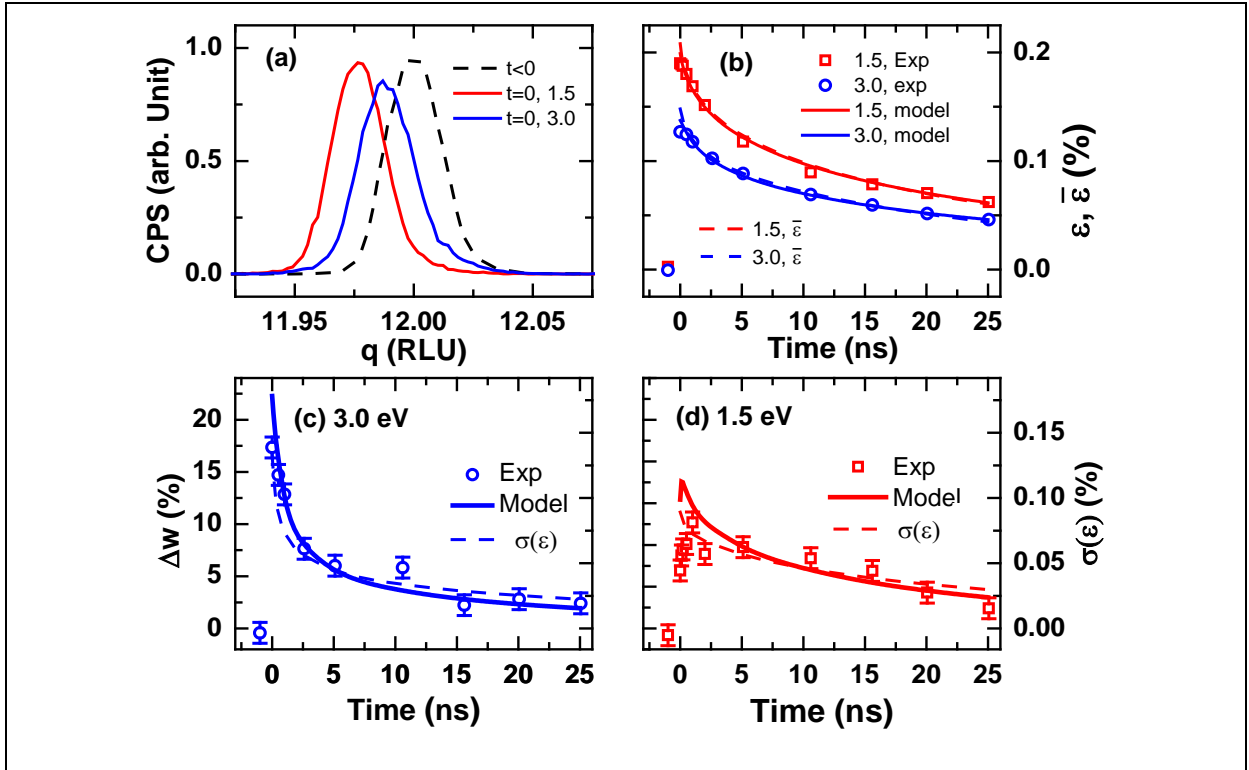

Figure. 1 (a) Shift of the (0 0 12) diffraction peak due to photon excitation at 1.5 and 3.0 eV photon energy for a 100 nm SIO film. (b) The strain $\varepsilon$ as a function of time for 1.5 and 3.0 eV photon energy. (c) and (d): fractional root mean square (rms) broadening ($\Delta w$) of the diffraction peak for the two excitation photon energies. Symbols are experimental data and the solid lines in (b-d) are fitting using the model described by Eq. (1-3). We also give the average strain ($\bar{\varepsilon}$) and the standard deviation ($\sigma(\varepsilon)$, right axis in (c, d)) of the modeled strain/phase profile using dashed lines in (b-d). The incident photon fluence is 20 mJ/cm$^2$. Note that, the shift of the diffraction peak represents the average phase and the broadening of the diffraction peak represents the dephasing of the diffracting unit cells in the film.

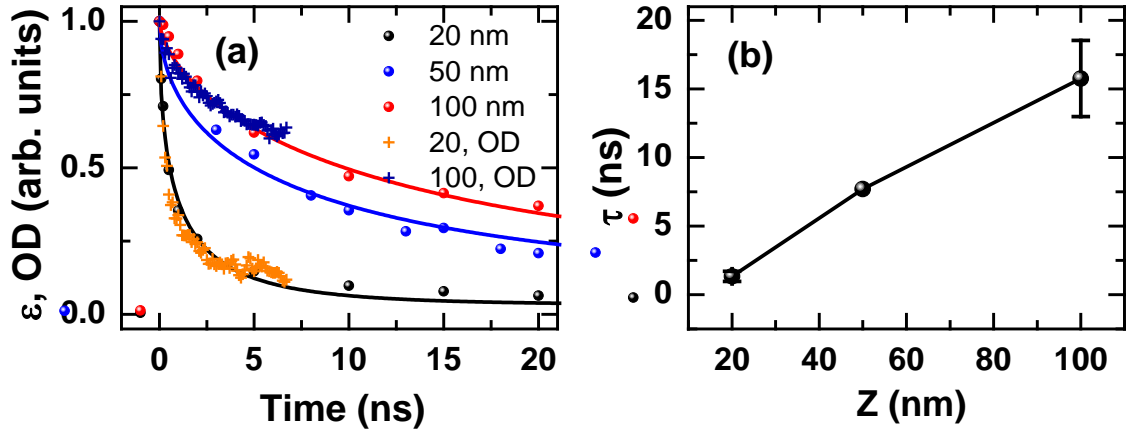

Figure 2 (a) The normalized strain recovery dynamics for 20, 50, and 100 nm films with 1.5 eV excitation. The corresponding excitation fluences are 20 mJ, 3.7, 20 mJ/cm$^2$, and the peak strains are 0.34%, 0.12%, and 0.19%, respectively. Also shown in (a) is the optical density (OD) from the TAS measurement for the 20 and 100 nm films with temporal resolution scaled down to 100 ps, more details in Fig. (S3). (b) The 1/e recovery time as a function of the film thickness, the error bars are due to the measurement for the films under different fluences and temperatures (See Fig. 4 (a) for the 20 nm film case). Using $\tau=(Z/\pi)^2/D$, an effective diffusion parameter $D=60$ nm$^2$ ns is obtained.

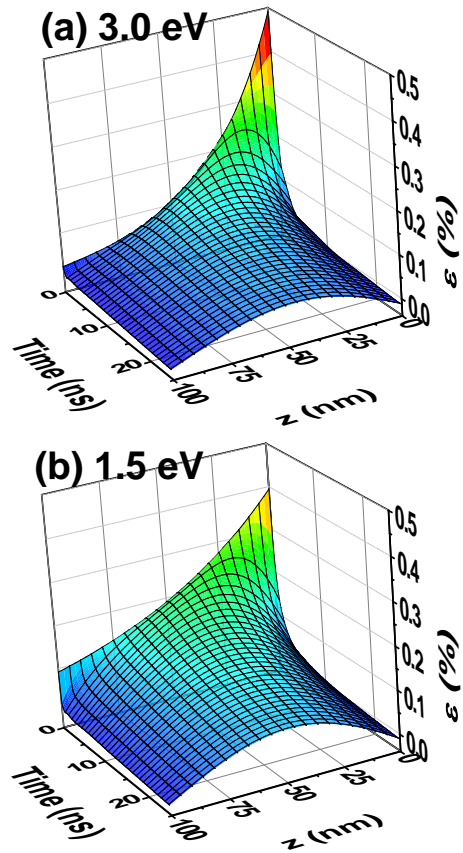

Figure 3. Reconstructed time-dependent strain map of the photo-induced strain using the diffusion model for the 100 nm film at (a) 3.0 eV and (b) 1.5 eV excitation photon energies. The strain is proportional to the local excitonic quasiparticle density. The difference between the two cases derives from the different initial photon deposition depths at time zero. Effectively, (a, b) are the spatiotemporal evolution of the coherence of the diffracting unit cells. The Fitting to the strain (average phase) and the broadening (de-coherence of the phase) of the diffracting unit cells are shown in Fig. 1.

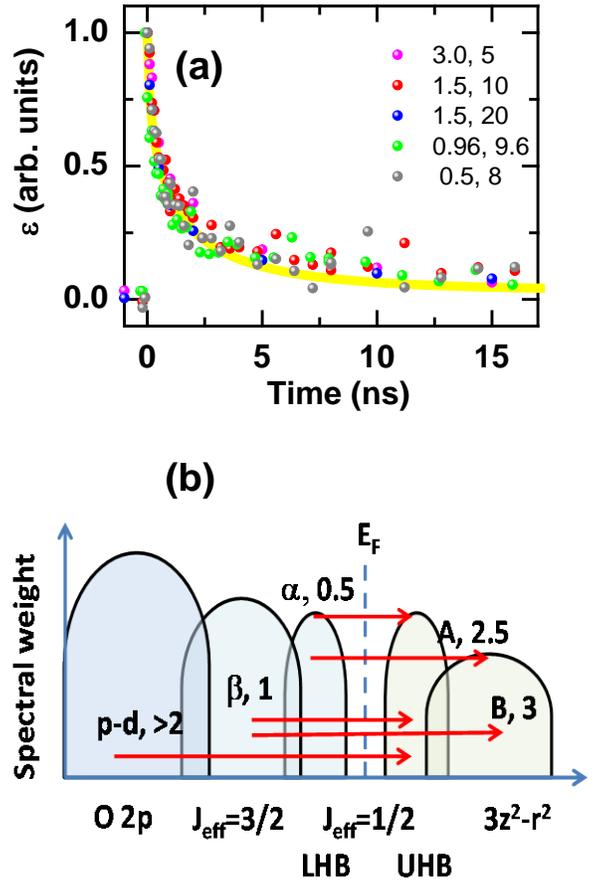

Figure 4. (b) The normalized recovery dynamics for the 20 nm film with excitation photon energy ranging from 0.5 to 30 eV. The excitation photon energy in eV and the fluence in mJ/cm$^2$ are indicated, the corresponding peak strains are 0.05%, 0.23%, 0.34%, 0.18%, and 0.07%. (b) Schematic band structure of SIO showing the different optical excitation pathways with corresponding excitation energy in eV.

# Lattice correlation of Hubbard excitons in a Mott insulator Sr2IrO4 and reconstruction of their hopping dynamics via time-dependent coherence analysis of the Bragg diffract

Yuelin Li[1*], Richard Schaller[2], Mengze Zhu[3], Donald A. Walko[1], Jungho Kim[1], Xianglin Ke[3], Ludi Miao[4], and Z. Q. Mao[4]

[1]Advanced Photon Source, Argonne National Laboratory, Argonne, Illinois 60439, USA

[2]Center of Nanoscale Materials, Argonne National Laboratory, Argonne, Illinois 60439, USA

[3]Department of Physics and Astronomy, Michigan State University, East Lansing, MI 48824, USA

[4]Department of Physics and Engineering Physics, Tulane University, New Orleans, LA 70118, USA

---

* ylli@aps.anl.gov

**Supplementary information**

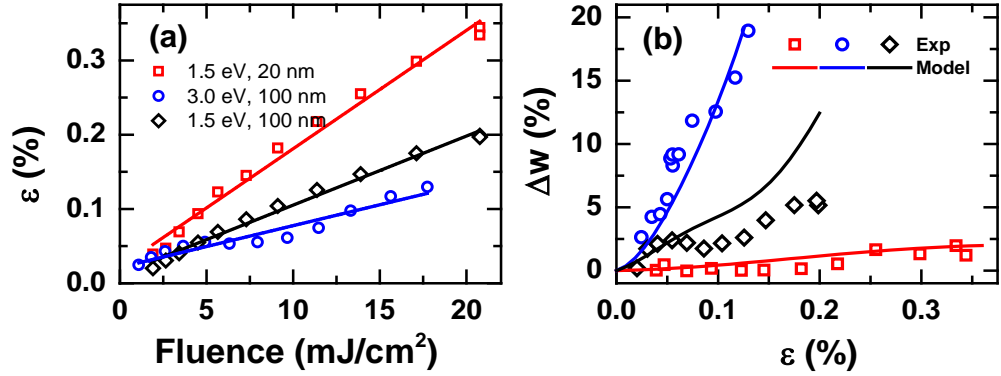

Figure S1 (a) Peak strain as a function of pump fluence under conditions indicated by the legend. The symbols are experiment data, the lines are linear fitting. (b) Fractional broadening of the diffraction peak at time zero as a function of the peak strain. In (b), symbols are experimental data and the lines are calculated using the model described by Eq. (1-3). The unperturbed rms width of the diffraction peak is 0.012 and 0.089 RLU for the 100 and 20 nm films, respectively.

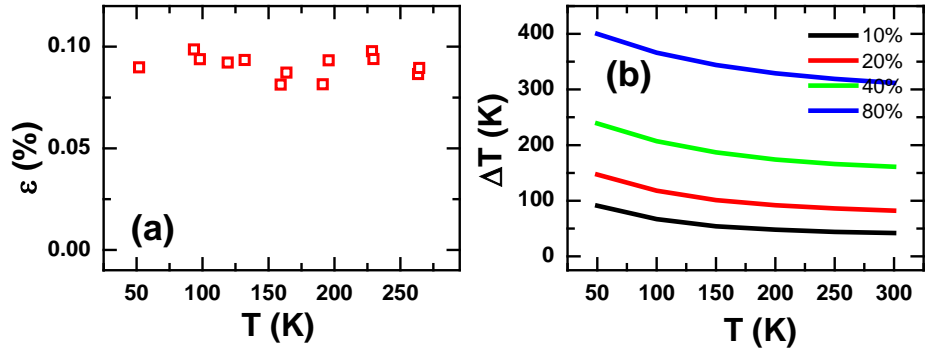

Figure S2 (a) Photo induced strain as a function of the sample temperature for the 20 nm film at a laser fluence of 3.7 mJ/cm$^2$ at 1.5 eV pump photon energy, showing a constant strain as a function of sample temperature. (b) Temperature change due to laser heating using the temperature dependent specific heat as a function of temperature [1]. Different heating efficiency is considered for a d-d transition energy of 1.1 eV. A smaller strain is expected at higher sample temperature if the lattice expansion is dominated by thermal effect.

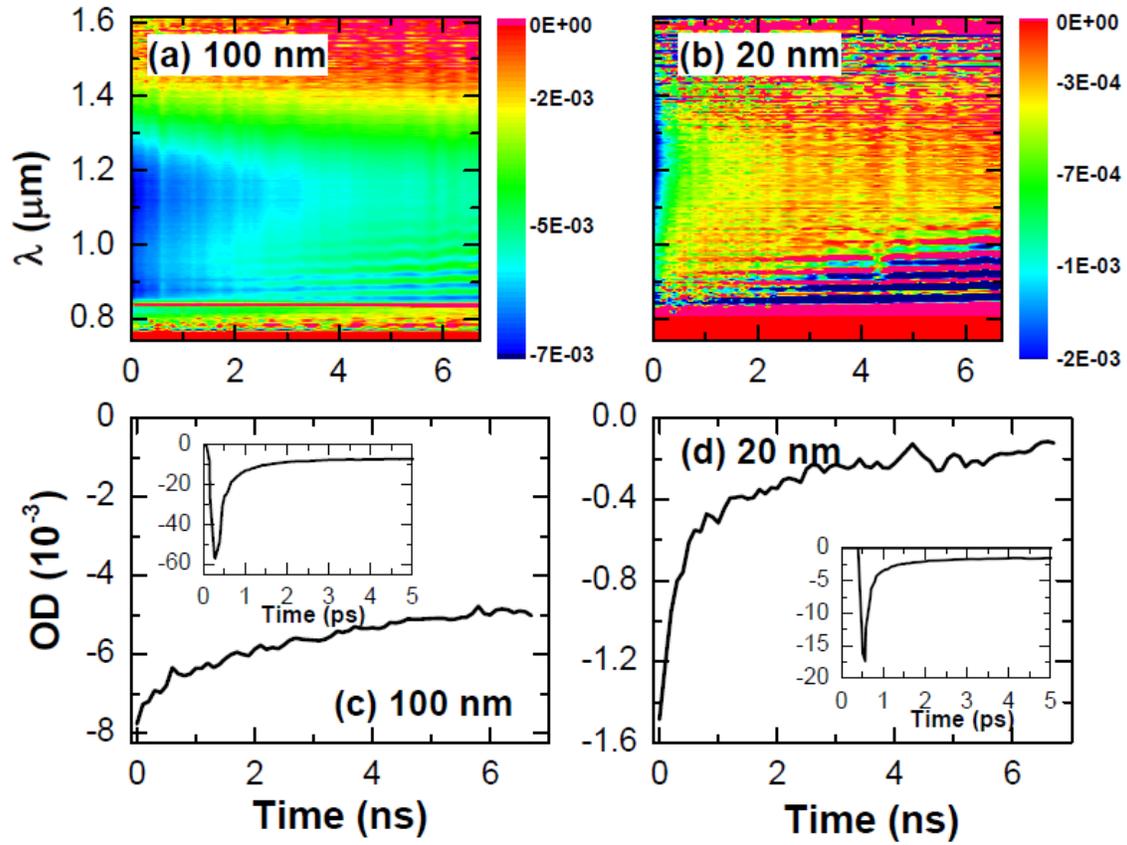

Figure. S3 (a, b) Transient absorption spectra of the 1.5 eV beam pumped 100 and 20 nm films and (c, d) the sum of the signal between 1.1 to 1.3 μm (0.9-0.95 eV) scaled to the 100 ps resolution for 100 nm. The short-time signal is shown in the inserts. The negative OD is the result of the pump induced transparency, which reflects the change of the electronic structure. The OD dynamics has a slow and a fast component. The fast component, responsible for about 80% of the OD recovery, decays in less than 1 ps, (inserts in (c, d)), and can be attributed to recombination and cooling of photo-excited singly- and doubly-occupied sites via hot magnon and phonon emission [2]. The slow component overlaps with the lattice dynamics (see Fig. 2). The horizontal striation in (a) and (b) are experiment artifacts.

Table S1 Parameter used and fitted for Eq. (1-3) to reproduce the data in Fig. 1

| Pump photon energy | 3.0 eV | 1.5 eV |
| --- | --- | --- |
| $\alpha$ | 29 nm | 70 nm |
| $D$ | 120 nm$^2$ ns$^{\gamma-1}$ | 120 nm$^2$ ns$^{\gamma-1}$ |
| $\gamma$ | 0.48 | 0.44 |
| $s(z)/D(t)$, $z=0$ | 12.5 nm$^{-1}$ | 12.5 nm$^{-1}$ |
| $s(z)/D(t)$, $z=Z$ | 154 nm$^{-1}$ | 154 nm$^{-1}$ |